\documentclass[twocolumn,epjc3,final]{svjour3}  
\smartqed  % flush right qed marks, e.g. at end of proof
\RequirePackage{graphicx}
\RequirePackage{mathptmx}      % use Times fonts if available on your TeX system
\RequirePackage{latexsym,amssymb,amsmath}
\RequirePackage[colorlinks,citecolor=blue,urlcolor=blue,linkcolor=blue]{hyperref}
\journalname{Eur. Phys. J. C}

\def\prl{Physical Review Letters}
\def\prd{Physical Review D}
\def\aap{Astron. Astroph.}
\def\apj{Astroph. J. }

\def\mnras{Mon. Not. R. Astron. Soc.}

\begin{document}

\title{Extracting the energy and angular momentum of a Kerr black hole%\thanksref{t1}
}
%\subtitle{Do you have a subtitle?\\ If so, write it here}

%\titlerunning{Short form of title}        % if too long for running head

\author{J. A. Rueda\thanksref{e1,addr1,addr2,addr3,addr4,addr5} \and R. Ruffini\thanksref{e2,addr1,addr2,addr6} %etc.
}

\thankstext{e1}{e-mail: jorge.rueda@icra.it}
\thankstext{e2}{e-mail: ruffini@icra.it}

\institute{ICRANet, Piazza della Repubblica 10, I--65122 Pescara, Italy \label{addr1}
           \and
           ICRA, Dipartimento di Fisica, Sapienza Universit\`a di Roma, P.le Aldo Moro 5, I--00185 Rome, Italy \label{addr2}
           \and
           ICRANet-Ferrara, Dipartimento di Fisica e Scienze della Terra, Universit\`a degli Studi di Ferrara, Via Saragat 1, I--44122 Ferrara, Italy \label{addr3}
           \and
           Dipartimento di Fisica e Scienze della Terra, Universit\`a degli Studi di Ferrara, Via Saragat 1, I--44122 Ferrara, Italy \label{addr4}
           \and
           INAF, Istituto de Astrofisica e Planetologia Spaziali, Via Fosso del Cavaliere 100, 00133 Rome, Italy \label{addr5}
           \and
           INAF, Viale del Parco Mellini 84, 00136 Rome  Italy\label{addr6}
}

\date{Received: date / Accepted: date}
% The correct dates will be entered by the editor

\maketitle

\begin{abstract}
It has been thought for decades that rotating black holes (BHs) power the energetic gamma-ray bursts (GRBs) and active galactic nuclei (AGNs), but the mechanism that extracts the BH energy has remained elusive. We here show that the solution to this problem arises when the BH is immersed in an external magnetic field and ionized low-density matter. For a magnetic field parallel to the BH spin, the induced electric field accelerates electrons outward and protons inward {in a conical region, centered on the BH rotation axis, and of semi-aperture angle $\theta \approx 60^\circ$ from the BH rotation axis}. For an antiparallel magnetic field, protons and electrons exchange their roles. The particles that are accelerated outward radiate off energy and angular momentum to infinity. The BH powers the process by reducing its energy and angular momentum by capturing polar protons and equatorial electrons with net negative energy and angular momentum. The electric potential allows for negative energy states outside the BH ergosphere, so the latter does not play any role in this electrodynamical BH energy extraction process.
\end{abstract}

%display desired
\maketitle

%\correspondingauthor{Liang Li}
%\email{}

%%%%%%%%%%%%%%%%%%%%%%%%%%%%%%%%%%%%%%%%%%%%%%%%%%%%
%%%%%%%%%%%%%%%%%%%%%%%%%%%%%%%%%%%%%%%%%%%%%%%%%%%%
\section{Introduction} \label{sec:1}
%%%%%%%%%%%%%%%%%%%%%%%%%%%%%%%%%%%%%%%%%%%%%%%%%%%%
%%%%%%%%%%%%%%%%%%%%%%%%%%%%%%%%%%%%%%%%%%%%%%%%%%%%

In this article, we show an electrodynamical process that efficiently extracts the rotational energy of BHs. The mechanism works for stellar-mass BHs in strong magnetic fields that power GRBs and supermassive BHs in weak magnetic fields that power AGNs. A critical ingredient for this discussion is one of the most relevant concepts of BHs, the Christodoulou-Ruffini-Hawking mass-energy formula \cite{1970PhRvL..25.1596C,1971PhRvD...4.3552C,1971PhRvL..26.1344H}. In its most general form, for a charged, rotating BH, it reads\footnote{We use geometric units $c=G=1$ unless otherwise specified.}
\begin{equation}\label{eq:Mbh}
M^2 = \left(M_{\rm irr} + \frac{Q^2}{4 M_{\rm irr}}\right)^2 + \frac{J^2}{4 M^2_{\rm irr}},
\end{equation}
which relates the BH mass-energy, $M$, to three independent pieces, the \textit{irreducible} mass, $M_{\rm irr}$, the charge, $Q$, and the angular momentum, $J$. The radius of the BH horizon is $r_H = M + \sqrt{M^2 - a^2 - Q^2}$, being $a=J/M$, the angular momentum per unit mass. Equation (\ref{eq:Mbh}) implies a great corollary: part of the BH energy is extractable, i.e., $E_{\rm ext} = M - M_{\rm irr}\geq 0$, and it amounts up to $50\%$ of the mass-energy of a non-rotating, charged BH (in the extreme case $Q=M$), and up to $29\%$ in a neutral, rotating BH (in the extreme case $a=M$). It is worth noticing that the above percentages are obtained under the nontrivial assumption that the BH irreducible mass remains constant during the energy extraction process. For fifty years as of this writing, the concept of BHs being energy storehouses usable by nature has permeated relativistic astrophysics at the theoretical and experimental levels.

To explain the most powerful transients in the Universe, GRBs, stellar-mass (i.e., of a few $M_\odot$) BHs should release up to a few $10^{54}$ erg in a few seconds. The supermassive BHs (of up to $10^9 M_\odot)$, to power AGNs, release luminosities of up to $10^{46}$~erg~s$^{-1}$ for billion years. Existing models of AGNs attempt to explain the emission with massive jets powered by an accretion disk around the BH, and most GRB models have inherited the same idea (see, e.g., \cite{2004RvMP...76.1143P,2018pgrb.book.....Z}, and references therein). Accretion disk models use gravitational energy, whose low efficiency makes it costly to power the most energetic processes in these relativistic sources.

The binary-driven hypernova (BdHN) model of GRBs has proposed as \textit{inner engine} of the high-energy emission in the gigaelectronvolt (GeV) domain, a Kerr BH surrounded by low-density matter and a magnetic field, modeled by the Wald solution \cite{2019ApJ...886...82R,2021A&A...649A..75M,2022ApJ...929...56R}. For an aligned and parallel (to the BH spin axis) magnetic field, the induced electric field in the polar region accelerates electrons outwardly, reaching ultrarelativistic energies and emitting synchrotron and high-energy curvature radiation. In \cite{2020EPJC...80..300R,2021A&A...649A..75M}, the model has been applied with $M=4.4 M_\odot$, $a/M=0.4$, and $B_0 = 4 \times 10^{10}$ G to the energetic GRB 190114C, and extended to AGNs, e.g., for the supermassive BH in M87*, with $M=6\times 10^9 M_\odot$, $a/M=0.1$, and $B_0 = 10$ G. These works have focused on the emission of escaping particles assuming by energy conservation that the Kerr BH pays for the energy radiated to infinity. On this basis, the evolution of the BH mass, angular momentum, and irreducible mass as the system radiates have been determined (see, e.g., \cite{2021A&A...649A..75M,2021MNRAS.504.5301R}). However, the mechanism for which the BH loses energy and angular momentum has remained unexplained.

Thus, extracting the BH energy is tantalizing and crucial in relativistic astrophysics. The first mechanism of BH energy extraction was the mechanical Penrose process \cite{1969NCimR...1..252P}. A particle of energy $E_1$ splits into two particles of energy $E_2$ and $E_3$ that, by energy conservation, fulfill $E_3 = E_1 - E_2$. So, $E_3 > E_1$ if $E_2 <0$, and the BH reduces its mass by $\delta M = E_2 <0$ and angular momentum by $\delta J = L_2 <0$ by absorbing such a particle. The split must occur in the BH \textit{ergosphere}, where negative energy (and associated negative angular momentum) states exist. We shall see that the ergosphere does not play any role in the electrodynamical mechanism presented here. 

It was soon demonstrated that the Penrose process is either unrealizable or inefficient (see, e.g., \cite{1971NPhS..229..177P,1972ApJ...178..347B,1974ApJ...191..231W}). Thus, numerous works have searched for alternatives. For example, from the mechanical viewpoint, the \textit{collisional} Penrose process has received much attention (see, e.g., \cite{2009PhRvL.103k1102B,2012PhRvL.109l1101B,2014PhRvL.113z1102S,2015PhRvL.114y1103B}). Generalizations of the Penrose process, i.e., the same three-body problem, accounting for electromagnetic fields, can be found, e.g., in \cite{1984PhRvD..30.1625D,1986ApJ...307...38P,2019Univ....5..125T}, and references therein. In parallel, increasing research has been devoted to electromagnetic fields to extract the BH energy. The idea of matter-dominated plasma accreting onto a Kerr BH by Ruffini and Wilson \cite{1975PhRvD..12.2959R}, further developed in the Blandford-Znajek mechanism \cite{1977MNRAS.179..433B}, which proposes that poloidal and toroidal magnetic field lines threading the BH extract its rotational energy. Without entering into the discussion of whether or not such a mechanism can operate in accreting rotating BHs, its efficiency, and power should be at most (although unlikely) that of the surrounding accretion disk \cite{1997MNRAS.292..887G,1999ApJ...512..100L}. Numerical, relativistic magnetohydrodynamics and particle-in-cell simulations have also studied the problem (e.g., \cite{2004MNRAS.350..427K,2005MNRAS.359..801K,2019PhRvL.122c5101P}). In the above literature, it is assumed (or achieved under specific conditions) that the density of charged particles in the magnetosphere is high enough to shorten any electric field so that \textit{force-free} electrodynamics applies. Those magnetospheres fulfill magnetic dominance, i.e., $B^2-E^2 >0$, and lack accelerating electric fields, i.e.,  $\mathbf{E}\cdot \mathbf{B} = 0$ everywhere. Those systems can not accelerate charged particles and emit radiation. To alleviate this drawback, it has been borrowed from pulsar theory the concept of \textit{gaps} \cite{1971ApJ...164..529S,1975ApJ...196...51R}, limited regions in the magnetosphere where the force-free condition is violated, {leading to regions where} $\mathbf{E}\cdot \mathbf{B}\neq 0$ (e.g., \cite{1977MNRAS.179..433B,2004MNRAS.350..427K}).

Most numerical simulations of BH magnetospheres use as initial condition the Wald solution of the Einstein-Maxwell equations \cite{1974PhRvD..10.1680W}, which describes a Kerr BH immersed in a test magnetic field, asymptotically uniform and aligned (parallel or antiparallel) to the BH spin. The Wald solution contains large regions where $\mathbf{E}\cdot \mathbf{B}\neq 0$ (see next section below). The force-free condition is achieved if the charge density in the magnetosphere exceeds the Goldreich-Julian value \cite{1969ApJ...157..869G}, $n_{\rm GJ}= \Omega B_0/(2 \pi c\,e)$, where $\Omega$ is the angular velocity of the corotating magnetic field lines. Since the BH angular velocity is $\Omega_H = a/(2 M r_H)$, for $M=4 M_\odot$, $a/M = 1$, $B_0 = 10^{13}$ G, we obtain $\rho_{\rm GJ} = m_p n_{\rm GJ}\approx 5 \times 10^{-9}$ g cm$^{-3}$. Although it looks like a very small value easy to exceed, numerical simulations show that, e.g., the matter density around the BH formed from the gravitational collapse of a neutron star in a BdHN can be as low as $\rho \sim 10^{-14}$ g cm$^{-3}$ \cite{2019ApJ...883..191R,2019ApJ...871...14B}, and the matter to electromagnetic energy density ratio as low as $8 \pi \rho/B_0^2\sim 10^{-10}$. These physical conditions are far from the ones explored in numerical simulations of screening plasma leading to force-free conditions starting from the Wald solution (see, e.g., \cite{2005MNRAS.359..801K,2019PhRvL.122c5101P}, in which both quantities have much higher values).

Bearing the above in mind, we hold on to the Wald solution and show a process that occurs in its electromagnetic field configuration, where $\mathbf{E}\cdot \mathbf{B}\neq 0$, {allowing the Kerr BH rotational energy extraction. This overcomes the original difficulty brought by the condition $\mathbf{E}\cdot \mathbf{B}= 0$ in Ruffini and Wilson \cite{1975PhRvD..12.2959R} and Blandford and Znajek \cite{1977MNRAS.179..433B} treatments.}

%%%%%%%%%%%%%%%%%%%%%%%%%%%%%%%%%%%%%%%%%%%%%%%%%%%%
\section{The electromagnetic field} \label{sec:2}
%%%%%%%%%%%%%%%%%%%%%%%%%%%%%%%%%%%%%%%%%%%%%%%%%%%%

In spheroidal Boyer-Lindquist coordinates $(t, r, \theta, \phi)$, the Kerr BH metric reads \cite{1968PhRv..174.1559C}\footnote{We use geometric units $c=G=1$ unless otherwise specified.}
\begin{align}\label{eq:metric}
    ds^2 &= -\left(1-\frac{2 M r}{\Sigma}\right) dt^2 + \frac{\Sigma}{\Delta} dr^2 + \Sigma\,d\theta^2 + \frac{A}{\Sigma}\sin^2\theta\,d\phi^2 \nonumber \\
    &- \frac{4 a M r}{\Sigma} \sin^2\theta\, dt\, d\phi,
\end{align}
where $\Sigma=r^2+a^2\cos^2\theta$, $\Delta=r^2-2 M r+ a^2$, $A = (r^2+a^2)^2-\Delta a^2 \sin^2\theta$, being $M$ and $a=J/M$, respectively, the BH mass and angular momentum per unit mass. 

The electromagnetic four-potential of the Wald solution for an uncharged, rotating BH is given by \cite{1974PhRvD..10.1680W}
\begin{equation}
    A_\mu = \frac{B_0}{2} \,\psi_\mu + a \,B_0\, \eta_\mu,
\end{equation}
where $B_0$ is the asymptotic magnetic field strength, and $\eta^\mu= \delta^\mu_{\hphantom{\mu}{t}}$ and $\psi^\mu=\delta^\mu_{\hphantom{\mu}{\phi}}$ are the time-like and space-like Killing vectors of the Kerr metric. In the frame of the \textit{locally non-rotating} (LNR) observer \cite{1970ApJ...162...71B,1972ApJ...178..347B}, which carries a tetrad basis with vectors $\vec{e}_{\hat a}$, the electric and magnetic field components are given by 
\begin{equation}
    E_{\hat{i}} = E_\mu\,\vec{e}^\mu_{\hphantom{\mu}{\hat i}} = F_{\hat{i} \hat{t}}$,\qquad $B_{\hat{i}} = B_\mu\,\vec{e}^\mu_{\hphantom{\mu}{\hat i}} = (1/2)\epsilon_{\hat{i}\hat{j}\hat{k}}F^{\hat{j}\hat{k}},
\end{equation}
where $F_{\mu \nu}$ is the electromagnetic field tensor in the coordinate basis. Careted components are in the LNR frame, Greek indexes run from $0$ to $3$ ($t$, $r$, $\theta$, and $\phi$), and Latin indexes run from $1$ to $3$. In Boyer-Lindquist coordinates, {the components $E_{\hat{i}}$ and $B_{\hat{i}}$ are given in Eqs. (16a)--(16d) in \cite{1978PhRvD..17.1518D} for the chargeless case ($Q=0$) and can be written as}

\begin{subequations}
{
\begin{align}
E_{\hat{r}} &= -\frac{B_0 a M}{\Sigma^2 A^{1/2}} \Bigg[(r^2 + a^2)(r^2-a^2\cos^2\theta)(1+\cos^2\theta) \nonumber \\
&- 2 r^2 \sin^2\theta\,\Sigma\Bigg],\label{eq:Er}\\
E_{\hat{\theta}} &= B_0 a M\,\frac{\Delta^{1/2}}{\Sigma^2 A^{1/2}} 2 r a^2 \sin\theta \cos\theta (1+\cos^2\theta),\label{eq:Etheta}
\end{align}
}
\end{subequations}
\begin{subequations}
{
\begin{align}
B_{\hat{r}} &= -\frac{B_0 \cos\theta}{\Sigma^2 A^{1/2}} \Bigg\{2 M r a^2 [2 r^2 \cos^2\theta+a^2(1+\cos^4\theta)] \nonumber \\
&-(r^2+a^2)\Sigma^2\Bigg\},\label{eq:Br}\\
B_{\hat{\theta}}&= -\frac{\Delta^{1/2}B_0 \sin\theta}{\Sigma^2 A^{1/2}} [M a^2 (r^2-a^2\cos^2\theta)(1+\cos^2\theta) \nonumber \\
&+ r \Sigma^2].\label{eq:Btheta}
\end{align}
}
\end{subequations}

{We can now calculate the regions of charged particles' acceleration. For the present case of magnetic dominance, i.e., $B^2>E^2$ \cite{1978PhRvD..17.1518D}, charged particles move along magnetic field lines. For magnetic field lines that cross the horizon, a particle will either move inward to the BH or be expelled outward, depending upon its charge. The electric field component parallel to the magnetic field line accelerates the particle. Therefore, we calculate the scalar product $\vec{E}\cdot \vec{B}$ on the BH horizon, $(\vec{E}\cdot \vec{B})_H$. Specifically, we are interested in the regions where $(\vec{E}\cdot \vec{B})_H$ is positive and negative, so we calculate the regions where it vanishes, which separate the regions of acceleration. At the event horizon, we have $\Delta = 0$, which leads to $(\vec{E}\cdot \vec{B})_H = (E_{\hat{r}}B_{\hat{r}})_H = E^H_{\hat{r}}B^H_{\hat{r}}$. Thus, the scalar product vanishes where either $E^H_{\hat{r}}$ or $B^H_{\hat{r}}$ vanishes. From Eq. (\ref{eq:Br}), we have $B^H_{\hat{r}} = 2 \cos^2\theta B_0 M r_H (r_H^2-a^2)/\Sigma_H^2$, which readily tells that $B^H_{\hat{r}} = 0$ on the equator, $\theta = \pi/2$. The solution of the equation $E^H_{\hat{r}} = 0$ is given by the angles $\theta_c$ that vanish the expression within the square brackets of Eq. (\ref{eq:Er}), i.e.,}
\begin{equation}\label{eq:thetac}
    \cos^2\theta_c = -\frac{\sigma}{2 a^2} + \frac{r_H}{a} \sqrt{1 + \left(\frac{\sigma}{2 a r_H}\right)^2},
\end{equation}
where $\sigma = (r_H^2-a^2)(r_H+M)/(r_H -M)$. {We recall that this is a spherical polar angle, so it is positively measured clockwise from the polar axis, and it is in the range $[0,\pi]$.} At second-order approximation in $a/M$, $\sigma \approx 2 r_H (r_H + M)$ and the above expression reduces to the one of \cite{1975PhRvD..12.3037K}, $\cos^2\theta_c \approx r_H/[2 (r_H + M)]$. At first order, $r_H \approx 2M$, so $\cos^2\theta_c \approx 1/3$. 
\begin{figure}
    \centering
    \includegraphics[width=\hsize,clip]{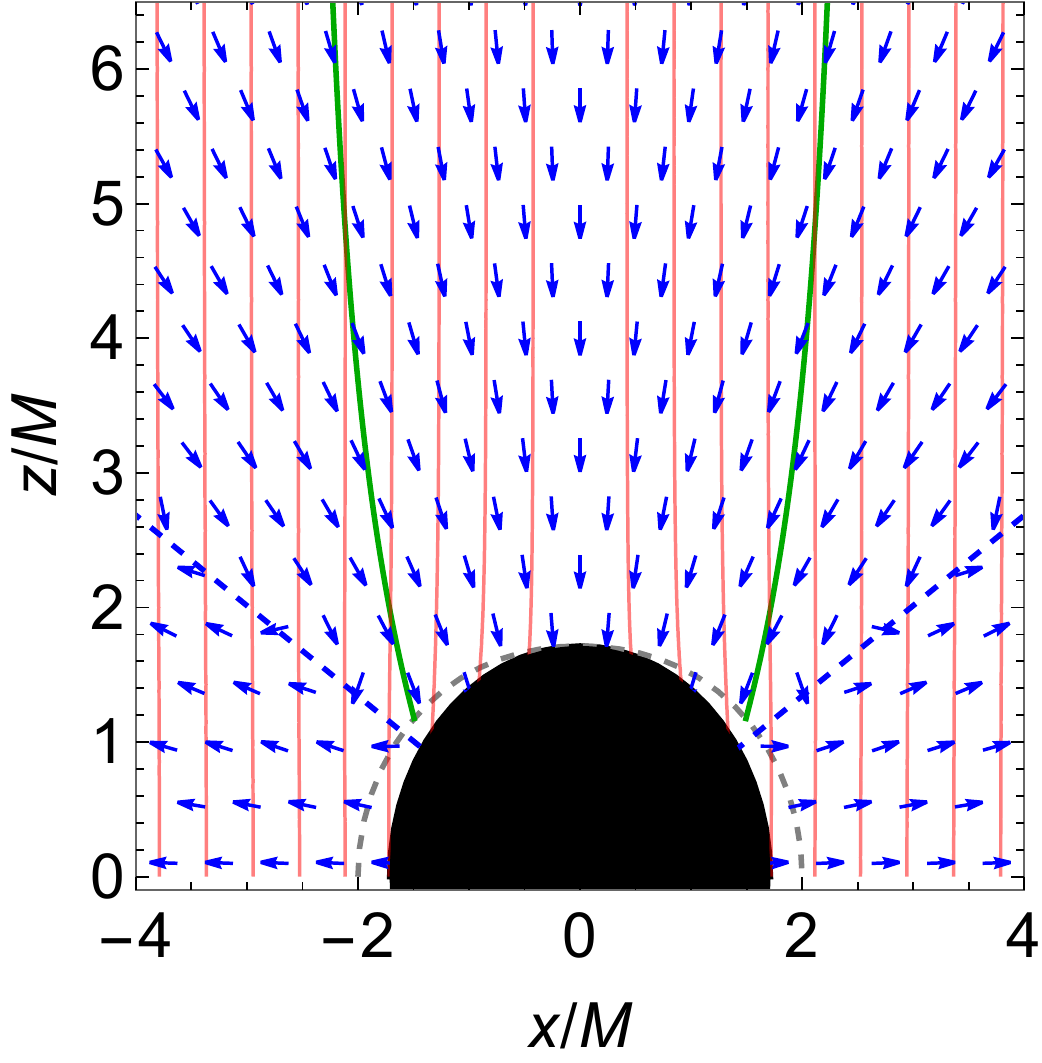}
    \caption{BH horizon (filled-black), ergosphere (dashed-gray), $K_p = 0$ boundary (green), electric field lines (blue arrows) and magnetic field lines (red, contours of constant $A_\phi$). The BH parameters are mass $M=4 M_\odot$, the spin parameter $a/M = 0.7$, and $B_0 = 4.4\times 10^9$ G. For the present spin parameter, $\theta_c\approx 56.12^\circ$, marked by the dashed-blue lines. The physical situation in the southern hemisphere is analogous due to equatorial symmetry. The figure shows the $xz$ plane ($\phi = 0, \pi$) in Cartesian Kerr-Schild coordinates (see, e.g., \cite{2022ApJ...929...56R}).}
    \label{fig:Kpzero}
\end{figure}

Figure \ref{fig:Kpzero} shows the electric field lines (blue arrows) and the magnetic field lines (contours of constant $A_\phi$, in red) for a BH with $a/M = 0.7$. {We display only the northern hemisphere for the azimuthal angles $\phi = 0$ and $\phi=\pi$. The physical situation is analogous in the southern hemisphere, given the equatorial reflection symmetry of the Wald solution. For this spin value, Eq. (\ref{eq:thetac}) leads to $\theta_{c,1}\approx 56.12^\circ$, shown by the dashed-blue line in the first quadrant (i.e., where $\phi = 0$, so $x>0$ and $z>0$). It also vanishes at $\theta_{c,2}\approx 123.88^\circ$ which lies in the fourth quadrant, not shown in the figure. Because of the axial symmetry, the scalar product also vanishes along a line given by the same $\theta_{c,1}$ and $\phi = \pi$, leading to the dashed-blue line in the second quadrant (i.e., where $x<0$ and $z>0$).}

{
Therefore, these blue-dashed lines separate four regions where $\vec{E}\cdot \vec{B} \neq 0$, unveiling the quadrupole nature of the electric field. We call hereafter as \textit{polar} the region within the two blue-dashed lines in the northern hemisphere. There is an analogous polar region in the southern hemisphere by equatorial reflection symmetry. We call as \textit{equatorial} the region $\theta_{c,1}\leq \theta \leq \theta_{c,2}$. There is an analogous equatorial region in the western hemisphere by axial symmetry.} 

The electric field is nearly radial in these regions. It decreases nearly as $1/r^2$, just like it would exist a net \textit{effective charge}  \cite{2019ApJ...886...82R,2021A&A...649A..75M}, $|Q_{\rm eff}| = 2 J B$. The net charge of the BH is zero, {as testified by calculating the induced charge over the horizon. For this task, one integrates the induced surface charge density introduced by Hanni and Ruffini \cite{1973PhRvD...8.3259H}, given by the discontinuity of the electric field component perpendicular to the BH horizon, i.e., the radial electric field. An explicit calculation for the Kerr BH immersed in the magnetic field can be found in \cite{1982MNRAS.198..339T,2000NCimB.115..751M,2022ApJ...929...56R}}. The induced charge on the two polar regions is of order $Q_{\rm eff}$ and is equal but of the opposite sign to the induced charge on the two equatorial regions \cite{2022ApJ...929...56R}. The concept of effective charge has been useful in the analysis of the high-energy (MeV and GeV) emission of GRBs in the BdHN model (see, e.g., \cite{2019ApJ...886...82R,2021PhRvD.104f3043M,2021A&A...649A..75M,2022EPJC...82..778R}). {The above effective charge is also known as the Wald charge, $Q_W$, derived in \cite{1974ApJ...191..231W} as the maximum charge the BH acquires by capturing charged particles along the polar axis, stopping accretion by the BH after it reaches $Q=Q_W$. We shall return to this point below in the conclusions.}

%%%%%%%%%%%%%%%%%%%%%%%%%%%%%%%%%%%%%%%%%%%%%%%%%%%%
%%%%%%%%%%%%%%%%%%%%%%%%%%%%%%%%%%%%%%%%%%%%%%%%%%%%
\section{Energy and angular momentum} \label{sec:3}
%%%%%%%%%%%%%%%%%%%%%%%%%%%%%%%%%%%%%%%%%%%%%%%%%%%%
%%%%%%%%%%%%%%%%%%%%%%%%%%%%%%%%%%%%%%%%%%%%%%%%%%%%

Therefore, we focus on capturing charged particles with negative energy and angular momentum. The conserved energy and angular momentum of charged particles are shifted by the presence of the electromagnetic potential so that negative energies are achievable well beyond the ergosphere, and co-rotating particles can attain negative angular momentum (details below). Interesting analyses of the motion properties of charged particles in the Wald solution can be found in \cite{2018PhRvD..98l3002L,2021PhRvD.104h4059G,2022MNRAS.512.2798K} (see also \cite{2021PhRvD.103b4021K} for the case of photons but in the ergosphere). 

The conserved energy and angular momentum of a particle of mass $m_i$ and charge $q_i$ are 
\begin{equation}\label{eq:EiLi}
    E_i = - \pi_\mu \eta^\mu = -\pi_0,\qquad L_i = \pi_\mu \psi^\mu = \pi_3,
\end{equation}
where $\pi_\alpha = p_\alpha + q_i A_\alpha$ is the canonical four-momentum, $p_\alpha = m_i u_\alpha$ the four-momentum, $u_\alpha$ the four-velocity, and $i = p,e$ stands for protons or electrons. Let us assume the particles are initially located at the position $(r_i, \theta_i, \phi_i)$, at rest. The latter condition implies that the particle lies initially outside the ergosphere, i.e., $\Sigma_{i} > 2 M r_{i}$, so $r_{i} > r_{\rm erg} = M + \sqrt{M^2-a^2 \cos^2\theta_{i}}$, and the initial four-velocity is $u^\alpha_{i} = u^0_{i} \delta^\alpha_0$, with $u^0_{i} = (1-2 M r_{i}/\Sigma_{i})^{-1/2}$. {From Eq. (\ref{eq:EiLi})}, the energy and angular momentum at the initial position are
\begin{subequations}\label{eq:EL}
\begin{align}
    E_i &= m_i \sqrt{1-\frac{2 M r_i}{\Sigma_i}} \pm e a B_0 \left[1 - \frac{M r_i}{\Sigma_{i}} (1+\cos^2\theta_{i})\right],\\
    L_{i} &= -m_i\frac{2 M a r_{i}\sin^2\theta_{i}}{\sqrt{\Sigma_{i} (\Sigma_{i}-2 M r_{i})}} \nonumber \\
    &\pm \frac{1}{2}e B_0 \sin^2\theta_{i} \left[ r^2_{i}+a^2 - \frac{2 M a^2 r_{i}}{\Sigma_{i}}(1 + \cos^2\theta_{i})  \right],
\end{align}
\end{subequations}
where $e$ is the fundamental charge, the upper ($+$) sign applies for protons and the lower ($-$) sign for electrons. The terms due to the electromagnetic potential largely dominate in Eqs for astrophysical parameters. (\ref{eq:EL}). In the case $B_0 \gg 0.011 (M_\odot/M)(m_{i}/m_e)$ G, $e B_0 M \gg m_i$, so Eqs. (\ref{eq:EL}) lead to polar protons with $E_p>0$ and $L_p > 0$, and equatorial electrons with $E_e <0$ and $L_e <0$.

Those electrons' negative energy states are physically possible if i) they do not reach infinity and ii) a local observer measures positive kinetic energy. The first condition is automatically satisfied since equatorial electrons are accelerated inward. The four-velocity of a regular local observer at the horizon can be constructed by the linear combination of the spacetime Killing vectors \cite{1973blho.conf...57C}: $l^\mu = \eta^\mu + \Omega_H \psi^\mu$, being $\Omega_H = a/(2 M r_H)$ the BH angular velocity. Therefore, the kinetic energy this observer measures when the particles cross the event horizon is 
\begin{equation}\label{eq:Ki}
    K_i = -p_\mu l^\mu|_H = E_i -\Omega_H L_i.
\end{equation}
For electrons, $K_e > 0$ at any angle in the equatorial region. For polar protons, the condition $K_p\geq 0$ constrains their initial position $(r_p,\theta_p)$, i.e., for given $r_p$, the boundary $K_p = 0$ defines a maximum value of $\theta_p$, say $\theta_{K_p}$. The maximum value of this angle occurs at $r_p = r_{\rm erg}$, say $\theta_{K_p,\rm max}$. Figure \ref{fig:Kpzero} shows the boundary $K_p = 0$ (dashed-green curves) for a BH with spin parameter $a/M = 0.7$, for which $\theta_{K_p,\rm max} \approx 51.81^\circ$.

\begin{figure}
    \centering
    \includegraphics[width=\hsize,clip]{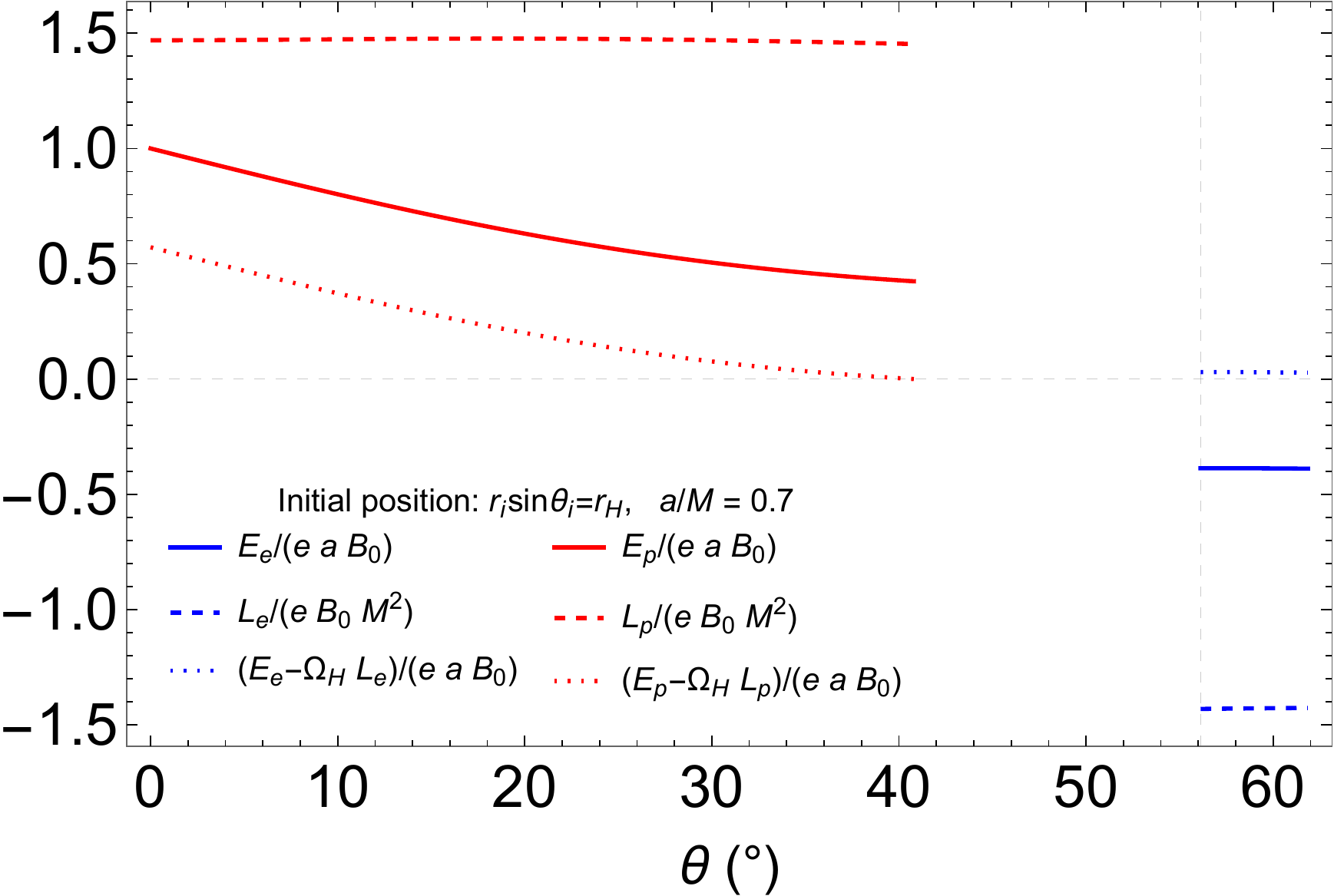}
    \caption{$E_i$, $L_i$, given by Eqs. (\ref{eq:EL}), and $E_i - \Omega_H L_i$, at initial positions outside the ergosphere ($r_i > r_{\rm erg}$) leading to the particle capture by the BH in the example of Fig. \ref{fig:Kpzero}.}
    \label{fig:EiLi}
\end{figure}

Charged particles will follow the magnetic field lines, and the latter point approximately in the $+z$ direction ($A_\phi = $ cons. implies $r\sin\theta \approx$ constant; see Fig. \ref{fig:Kpzero}), so the BH can capture those particles whose initial position fulfills $r_i\sin\theta_i \leq r_H$. Thus, the BH captures polar protons at $(r_p,\theta_p)$ within $0\leq \theta_p \leq \theta_{p, \rm max}$, where $\theta_{p, \rm max} = {\rm Min}(\theta_{K_p},\theta_{p, \rm cyl})$, $\theta_{p, \rm cyl} = \arcsin(r_H/r_p)$, and equatorial electrons at $(r_e,\theta_e)$ within $\theta_c \leq \theta_e \leq \theta_{e, \rm max}$, where $\theta_{e, \rm max} = \theta_{e, \rm cyl} = \arcsin(r_H/r_e)$. Figure \ref{fig:EiLi} shows $E_{p,e}$ and $L_{p,e}$ at initial positions that satisfy the capture conditions mentioned above, specifically for protons initially located in the polar region at $(r_p,\theta_p)$, with $r_p = r_H/\sin\theta_p$ and $0\leq \theta_p \leq \theta_{p,\rm max}$, and electrons in the equatorial region at $(r_e,\theta_e)$, with $r_e = r_H/\sin\theta_e$ and $\theta_c \leq \theta_e \leq \theta_{e,\rm max}$. For the present spin parameter, $a/M = 0.7$, the reference angles are $\theta_c \approx 56.12^\circ$, $\theta_{p, \rm max} = \theta_{p, \rm cyl}\approx 40.80^\circ$, and $\theta_{e,\rm max} = \theta_{e, \rm cyl}\approx 61.86^\circ$.

%%%%%%%%%%%%%%%%%%%%%%%%%%%%%%%%%%%%%%%%%%%%%%%%%%%%
%%%%%%%%%%%%%%%%%%%%%%%%%%%%%%%%%%%%%%%%%%%%%%%%%%%%
\section{Discussion and Conclusion} \label{sec:4}
%%%%%%%%%%%%%%%%%%%%%%%%%%%%%%%%%%%%%%%%%%%%%%%%%%%%
%%%%%%%%%%%%%%%%%%%%%%%%%%%%%%%%%%%%%%%%%%%%%%%%%%%%

{We have analyzed the capture of charged particles by a Kerr BH embedded in a test, asymptotically aligned magnetic field given by the Wald solution. Paper \cite{1974PhRvD..10.1680W} envisaged a situation in which the BH, by capturing charged particles along the rotation axis, gain charge up to a maximal possible value, $Q_W = 2 J B_0$. After that point, the charged particle accretion should stop. Our results show that the physical situation can be more complicated and interesting. To account for the feedback of the particle capture on the BH parameters and, as we discuss below, the distribution of particles, are essential to draw any conclusions on the BH evolution. Indeed, the charged particles' energy and angular momentum at different radii and latitudes can lead to a very different scenario. Second, special attention must be paid to estimating the change of all BH parameters in the process, including its irreducible mass. The latter is of paramount relevance to assess the efficiency and plausibility of the energy extraction process.}

\begin{figure}
    \centering
    \includegraphics[width=\hsize,clip]{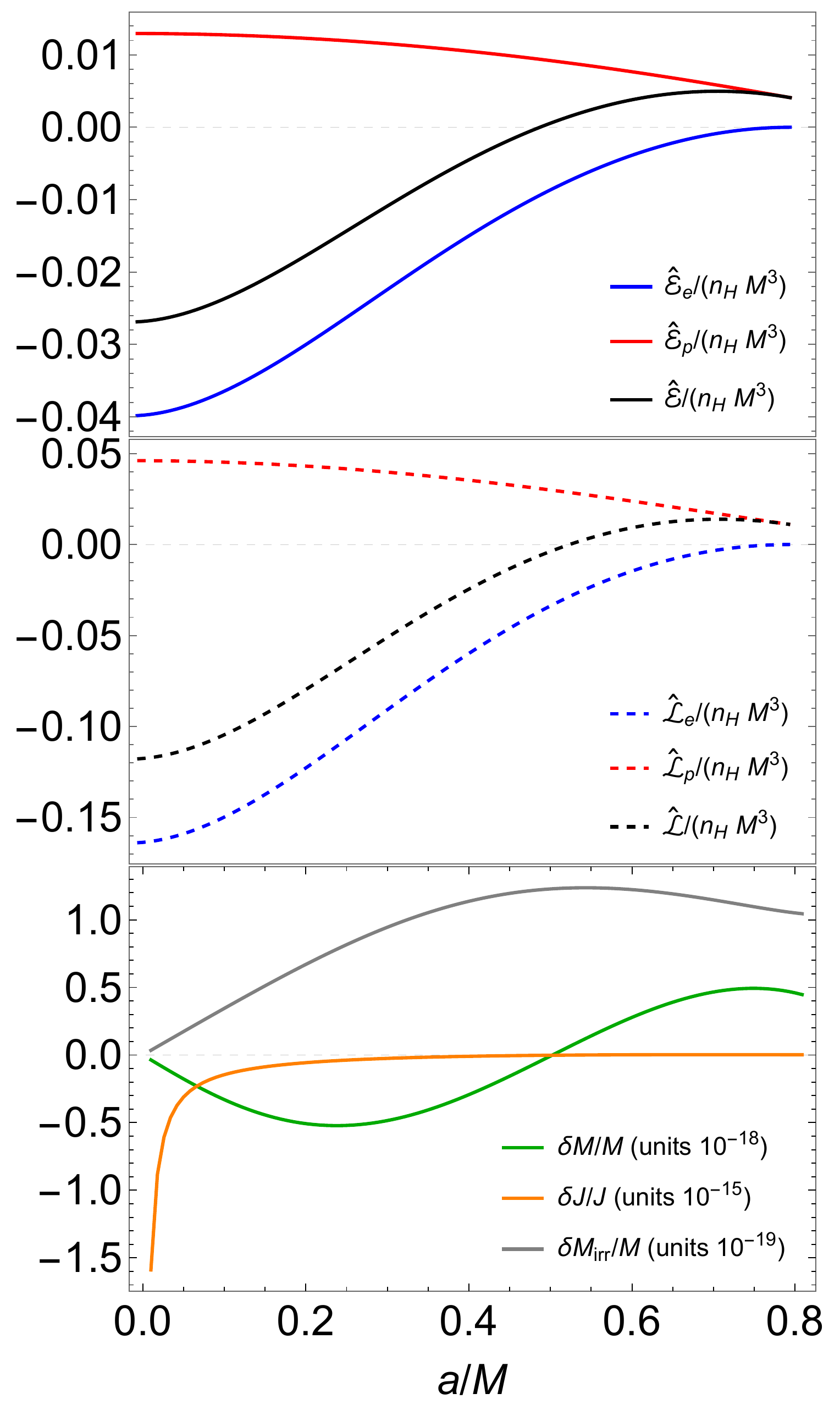}
    \caption{Upper: $\hat{{\cal E}}_i = {\cal E}_i/(e B_0 a)$ for polar protons (red) and equatorial electrons (blue) that cross the BH horizon. The net energy, $\hat{{\cal E}} = \hat{{\cal E}}_p+\hat{{\cal E}}_e$, is shown in black. Middle: analogous to the upper panel but for $\hat{{\cal L}}_i= {\cal L}_i/(e B_0 M^2)$, and $\hat{{\cal L}}$. Lower: Fractional change of the BH mass, $\delta M/M$ (green, units of $10^{-18}$), angular momentum, $\delta J/J$ (orange, units of $10^{-15}$), and irreducible mass, $\delta M_{\rm irr}/M$ (gray, units of $10^{-19}$). The particle density is $n(r,\theta) = {\cal N}(r)\Psi(\theta)$, where ${\cal N}(r) = n_H (r_H/r)^m$, with $m = 2$, $n_H = 6.0 \times 10^{10}$ cm$^{-3}$, and $\Psi(\theta) = (1-\cos\theta)^2$. In the upper and middle panels, the dimensionless energy and angular momentum are normalized by $n_H M^3$. This example uses $B_0 = B_c = 2\pi m_e^2 c^3/(e\,h) \approx 4.41 \times 10^{13}$ G.}
    \label{fig:ELtot}
\end{figure}

When the BH captures a proton or electron, its mass, angular momentum, and irreducible mass change by 
\begin{subequations}
    \begin{align}
        \delta M &= E_i,\\
        \delta J &= L_i,\\
        \delta M_{\rm irr} &= \frac{M_{\rm irr}}{\sqrt{M^2-a^2}}(\delta M - \Omega_H \delta J).
    \end{align}
\end{subequations}
Because $\delta M - \Omega_H \delta J = E_i - \Omega_H L_i = K_i \geq 0$ (see Fig. \ref{fig:EiLi}), we have $\delta M^2_{\rm irr} \geq 0$, as expected \cite{1970PhRvL..25.1596C,1971PhRvD...4.3552C,1971PhRvL..26.1344H}. Figure \ref{fig:EiLi} shows that the equatorial region from which the BH captures electrons is smaller than the polar region from which it captures protons. The main reason is that the magnetic field lines are parallel to the $z$-axis in the Wald solution, even in the BH vicinity. This is confirmed by the magnetic flux threading the BH horizon
\begin{equation}\label{eq:Phi}
    \Phi_B = \iint F_{23} d\theta d\phi,
\end{equation}
which leads to the ratio of the polar to equatorial flux 
\begin{equation}
    \frac{1+\sqrt{5}}{2} \leq \frac{\Phi_B(0,\theta_c)}{\Phi_B(\theta_c,\pi/2)}= \frac{r_H}{2 M} \tan^2\theta_c \leq 2,
\end{equation}
for $0\leq a/M \leq 1$. Thus, for the given magnetic field geometry, whether the net energy and angular momentum that the BH absorbs are negative or positive depending on the density of protons and electrons, $n$. Assuming local neutrality, protons, and electrons of number density $n$ transfer to the BH an energy 
\begin{subequations}
    \begin{align}
        {\cal E}_i &\approx 2 \pi \iint E_i n \sqrt{-g} dr d\theta,\\
        {\cal L}_i &\approx 2 \pi \iint L_i n \sqrt{-g} dr d\theta,
    \end{align}
\end{subequations}
where $g = - \Sigma^2 \sin^2\theta$ is the Kerr metric determinant. The constraints of the previous section give the integration boundaries. For a spherically symmetric density, $n = n(r)$, there are more capturable protons than electrons. The BH would acquire an energy ${\cal E} = {\cal E}_e + {\cal E}_p >0$ and angular momentum ${\cal L} = {\cal L}_e + {\cal L}_p >0$. An interesting situation occurs for an anisotropic density that increases towards the equator. As an example, Fig. \ref{fig:ELtot} shows ${\cal E}_e$, ${\cal E}_p$, ${\cal L}_e$, ${\cal L}_p$, ${\cal E}$ and ${\cal L}$, for $n(r,\theta) = {\cal N}(r)\Psi(\theta)$, where ${\cal N}(r) = n_H (r_H/r)^m$, and $\Psi(\theta) = (1-\cos\theta)^2$, with $m = 2$, $n_H = 6.0 \times 10^{10}$ cm$^{-3}$, which corresponds to a rest-mass density $\rho = 10^{-13}$ g cm$^{-3}$ near the BH horizon at the pole. We obtain ${\cal E}_p < |{\cal E}_e|$ and ${\cal L}_p < |{\cal L}_e|$, leading to ${\cal E}<0$ and ${\cal L}<0$, for values of the BH spin parameter $a/M\lesssim 0.5$.

%%%%%%%%%%%%%%%%%%%%%%%%%%%%%%%%%%%%%%%%%%%%%%%%%%%%
%%%%%%%%%%%%%%%%%%%%%%%%%%%%%%%%%%%%%%%%%%%%%%%%%%%%
%\section{Conclusions}\label{sec:5}
%%%%%%%%%%%%%%%%%%%%%%%%%%%%%%%%%%%%%%%%%%%%%%%%%%%%
%%%%%%%%%%%%%%%%%%%%%%%%%%%%%%%%%%%%%%%%%%%%%%%%%%%%

Therefore, the long-standing question of how to extract the rotational energy of a Kerr BH is answered naturally by analyzing a rotating BH capturing not a single charged particle but a bunch of them of opposite charges, at different latitudes (see Fig. \ref{fig:EiLi}). We have shown, using the Wald solution, that the electrodynamical extraction of rotational energy works for an anisotropic density of protons and electrons increasing with latitude (see Figs. \ref{fig:EiLi} and \ref{fig:ELtot}). Estimating the present rotational energy extraction process for different magnetic field configurations, matter accretion of varying nature, and more extended BH parameters, including non-vanishing electric charge, is now possible (Rueda and Ruffini, in preparation).
%  

%%%%%%%%%%%%%%%%%%%%%%%%%%%%%%%%%%%%%%%%%%%%%%%%%%%%%%%%
%\bibliography{references}

\begin{thebibliography}{10}
\providecommand{\url}[1]{{#1}}
\providecommand{\urlprefix}{URL }
\expandafter\ifx\csname urlstyle\endcsname\relax
  \providecommand{\doi}[1]{DOI \discretionary{}{}{}#1}\else
  \providecommand{\doi}{DOI \discretionary{}{}{}\begingroup
  \urlstyle{rm}\Url}\fi

\bibitem{1970PhRvL..25.1596C}
D.~{Christodoulou}, \prl \textbf{25}(22), 1596 (1970).
\newblock \doi{10.1103/PhysRevLett.25.1596}

\bibitem{1971PhRvD...4.3552C}
D.~{Christodoulou}, R.~{Ruffini}, \prd \textbf{4}, 3552 (1971).
\newblock \doi{10.1103/PhysRevD.4.3552}

\bibitem{1971PhRvL..26.1344H}
S.W. {Hawking}, Physical Review Letters \textbf{26}, 1344 (1971).
\newblock \doi{10.1103/PhysRevLett.26.1344}

\bibitem{2004RvMP...76.1143P}
T.~{Piran}, Reviews of Modern Physics \textbf{76}, 1143 (2004).
\newblock \doi{10.1103/RevModPhys.76.1143}

\bibitem{2018pgrb.book.....Z}
B.~{Zhang}, \emph{{The Physics of Gamma-Ray Bursts}} (Cambridge Univeristy
  Press, 2018).
\newblock \doi{10.1017/9781139226530}

\bibitem{2019ApJ...886...82R}
R.~{Ruffini}, R.~{Moradi}, J.A. {Rueda}, L.~{Becerra}, C.L. {Bianco},
  C.~{Cherubini}, S.~{Filippi}, Y.C. {Chen}, M.~{Karlica}, N.~{Sahakyan},
  Y.~{Wang}, S.S. {Xue}, \apj \textbf{886}(2), 82 (2019).
\newblock \doi{10.3847/1538-4357/ab4ce6}

\bibitem{2021A&A...649A..75M}
R.~{Moradi}, J.A. {Rueda}, R.~{Ruffini}, Y.~{Wang}, \aap \textbf{649}, A75
  (2021).
\newblock \doi{10.1051/0004-6361/201937135}

\bibitem{2022ApJ...929...56R}
J.A. {Rueda}, R.~{Ruffini}, R.P. {Kerr}, \apj \textbf{929}(1), 56 (2022).
\newblock \doi{10.3847/1538-4357/ac5b6e}

\bibitem{2020EPJC...80..300R}
J.A. {Rueda}, R.~{Ruffini}, European Physical Journal C \textbf{80}(4), 300
  (2020).
\newblock \doi{10.1140/epjc/s10052-020-7868-z}

\bibitem{2021MNRAS.504.5301R}
R.~{Ruffini}, R.~{Moradi}, J.A. {Rueda}, L.~{Li}, N.~{Sahakyan}, Y.C. {Chen},
  Y.~{Wang}, Y.~{Aimuratov}, L.~{Becerra}, C.L. {Bianco}, C.~{Cherubini},
  S.~{Filippi}, M.~{Karlica}, G.J. {Mathews}, M.~{Muccino}, G.B. {Pisani}, S.S.
  {Xue}, \mnras \textbf{504}(4), 5301 (2021).
\newblock \doi{10.1093/mnras/stab724}

\bibitem{1969NCimR...1..252P}
R.~{Penrose}, Nuovo Cimento Rivista Serie \textbf{1} (1969)

\bibitem{1971NPhS..229..177P}
R.~{Penrose}, R.M. {Floyd}, Nature Physical Science \textbf{229}(6), 177
  (1971).
\newblock \doi{10.1038/physci229177a0}

\bibitem{1972ApJ...178..347B}
J.M. {Bardeen}, W.H. {Press}, S.A. {Teukolsky}, \apj \textbf{178}, 347 (1972).
\newblock \doi{10.1086/151796}

\bibitem{1974ApJ...191..231W}
R.M. {Wald}, \apj \textbf{191}, 231 (1974).
\newblock \doi{10.1086/152959}

\bibitem{2009PhRvL.103k1102B}
M.~{Ba{\~n}ados}, J.~{Silk}, S.M. {West}, \prl \textbf{103}(11), 111102 (2009).
\newblock \doi{10.1103/PhysRevLett.103.111102}

\bibitem{2012PhRvL.109l1101B}
M.~{Bejger}, T.~{Piran}, M.~{Abramowicz}, F.~{H{\r{a}}kanson}, \prl
  \textbf{109}(12), 121101 (2012).
\newblock \doi{10.1103/PhysRevLett.109.121101}

\bibitem{2014PhRvL.113z1102S}
J.D. {Schnittman}, \prl \textbf{113}(26), 261102 (2014).
\newblock \doi{10.1103/PhysRevLett.113.261102}

\bibitem{2015PhRvL.114y1103B}
E.~{Berti}, R.~{Brito}, V.~{Cardoso}, \prl \textbf{114}(25), 251103 (2015).
\newblock \doi{10.1103/PhysRevLett.114.251103}

\bibitem{1984PhRvD..30.1625D}
S.V. {Dhurandhar}, N.~{Dadhich}, \prd \textbf{30}(8), 1625 (1984).
\newblock \doi{10.1103/PhysRevD.30.1625}

\bibitem{1986ApJ...307...38P}
S.~{Parthasarathy}, S.M. {Wagh}, S.V. {Dhurandhar}, N.~{Dadhich}, \apj
  \textbf{307}, 38 (1986).
\newblock \doi{10.1086/164390}

\bibitem{2019Univ....5..125T}
A.~{Tursunov}, N.~{Dadhich}, Universe \textbf{5}(5), 125 (2019).
\newblock \doi{10.3390/universe5050125}

\bibitem{1975PhRvD..12.2959R}
R.~{Ruffini}, J.R. {Wilson}, \prd \textbf{12}(10), 2959 (1975).
\newblock \doi{10.1103/PhysRevD.12.2959}

\bibitem{1977MNRAS.179..433B}
R.D. {Blandford}, R.L. {Znajek}, \mnras \textbf{179}, 433 (1977).
\newblock \doi{10.1093/mnras/179.3.433}

\bibitem{1997MNRAS.292..887G}
P.~{Ghosh}, M.A. {Abramowicz}, \mnras \textbf{292}(4), 887 (1997).
\newblock \doi{10.1093/mnras/292.4.887}

\bibitem{1999ApJ...512..100L}
M.~{Livio}, G.I. {Ogilvie}, J.E. {Pringle}, \apj \textbf{512}(1), 100 (1999).
\newblock \doi{10.1086/306777}

\bibitem{2004MNRAS.350..427K}
S.S. {Komissarov}, \mnras \textbf{350}(2), 427 (2004).
\newblock \doi{10.1111/j.1365-2966.2004.07598.x}

\bibitem{2005MNRAS.359..801K}
S.S. {Komissarov}, \mnras \textbf{359}(3), 801 (2005).
\newblock \doi{10.1111/j.1365-2966.2005.08974.x}

\bibitem{2019PhRvL.122c5101P}
K.~{Parfrey}, A.~{Philippov}, B.~{Cerutti}, \prl \textbf{122}(3), 035101
  (2019).
\newblock \doi{10.1103/PhysRevLett.122.035101}

\bibitem{1971ApJ...164..529S}
P.A. {Sturrock}, \apj \textbf{164}, 529 (1971).
\newblock \doi{10.1086/150865}

\bibitem{1975ApJ...196...51R}
M.A. {Ruderman}, P.G. {Sutherland}, \apj \textbf{196}, 51 (1975).
\newblock \doi{10.1086/153393}

\bibitem{1974PhRvD..10.1680W}
R.M. {Wald}, \prd \textbf{10}, 1680 (1974).
\newblock \doi{10.1103/PhysRevD.10.1680}

\bibitem{1969ApJ...157..869G}
P.~{Goldreich}, W.H. {Julian}, \apj \textbf{157}, 869 (1969).
\newblock \doi{10.1086/150119}

\bibitem{2019ApJ...883..191R}
R.~{Ruffini}, J.D. {Melon Fuksman}, G.V. {Vereshchagin}, \apj \textbf{883}(2),
  191 (2019).
\newblock \doi{10.3847/1538-4357/ab3c51}

\bibitem{2019ApJ...871...14B}
L.~{Becerra}, C.L. {Ellinger}, C.L. {Fryer}, J.A. {Rueda}, R.~{Ruffini}, \apj
  \textbf{871}(1), 14 (2019).
\newblock \doi{10.3847/1538-4357/aaf6b3}

\bibitem{1968PhRv..174.1559C}
B.~{Carter}, Physical Review \textbf{174}(5), 1559 (1968).
\newblock \doi{10.1103/PhysRev.174.1559}

\bibitem{1970ApJ...162...71B}
J.M. {Bardeen}, \apj \textbf{162}, 71 (1970).
\newblock \doi{10.1086/150635}

\bibitem{1978PhRvD..17.1518D}
T.~{Damour}, R.S. {Hanni}, R.~{Ruffini}, J.R. {Wilson}, \prd \textbf{17}(6),
  1518 (1978).
\newblock \doi{10.1103/PhysRevD.17.1518}

\bibitem{1975PhRvD..12.3037K}
A.R. {King}, J.P. {Lasota}, W.~{Kundt}, \prd \textbf{12}(10), 3037 (1975).
\newblock \doi{10.1103/PhysRevD.12.3037}

\bibitem{1973PhRvD...8.3259H}
R.S. {Hanni}, R.~{Ruffini}, \prd \textbf{8}(10), 3259 (1973).
\newblock \doi{10.1103/PhysRevD.8.3259}

\bibitem{1982MNRAS.198..339T}
K.S. {Thorne}, D.~{MacDonald}, \mnras \textbf{198}, 339 (1982).
\newblock \doi{10.1093/mnras/198.2.339}

\bibitem{2000NCimB.115..751M}
G.~{Miniutti}, R.~{Ruffini}, Nuovo Cimento B Serie \textbf{115}, 751 (2000)

\bibitem{2021PhRvD.104f3043M}
R.~{Moradi}, J.A. {Rueda}, R.~{Ruffini}, L.~{Li}, C.L. {Bianco}, S.~{Campion},
  C.~{Cherubini}, S.~{Filippi}, Y.~{Wang}, S.S. {Xue}, \prd \textbf{104}(6),
  063043 (2021).
\newblock \doi{10.1103/PhysRevD.104.063043}

\bibitem{2022EPJC...82..778R}
F.~{Rastegarnia}, R.~{Moradi}, J.A. {Rueda}, R.~{Ruffini}, L.~{Li},
  S.~{Eslamzadeh}, Y.~{Wang}, S.S. {Xue}, European Physical Journal C
  \textbf{82}(9), 778 (2022).
\newblock \doi{10.1140/epjc/s10052-022-10750-x}

\bibitem{2018PhRvD..98l3002L}
J.~{Levin}, D.J. {D'Orazio}, S.~{Garcia-Saenz}, \prd \textbf{98}(12), 123002
  (2018).
\newblock \doi{10.1103/PhysRevD.98.123002}

\bibitem{2021PhRvD.104h4059G}
K.~{Gupta}, Y.T.A. {Law}, J.~{Levin}, \prd \textbf{104}(8), 084059 (2021).
\newblock \doi{10.1103/PhysRevD.104.084059}

\bibitem{2022MNRAS.512.2798K}
S.S. {Komissarov}, \mnras \textbf{512}(2), 2798 (2022).
\newblock \doi{10.1093/mnras/stab2686}

\bibitem{2021PhRvD.103b4021K}
M.~{Kolo{\v{s}}}, A.~{Tursunov}, Z.~{Stuchl{\'\i}k}, \prd \textbf{103}(2),
  024021 (2021).
\newblock \doi{10.1103/PhysRevD.103.024021}

\bibitem{1973blho.conf...57C}
B.~{Carter}, in \emph{Black Holes (Les Astres Occlus)} (1973), pp. 57--214

\end{thebibliography}
%\bibliographystyle{spphys}

\end{document}